# INSTRUMENTATION FOR GEOLOGICAL FIELD WORK ON THE MOON

D. L. TALBOYS, G. W. FRASER, R. M. AMBROSI, N. NELMS
N. P. BANNISTER, M. R. SIMS, D. PULLAN and J. HOLT
*Space Research Centre, University of Leicester, University Road, Leicester, LE2 7RH UK*
*(E-mail: dlt3@star.le.ac.uk)*

**Abstract.** A human return to the Moon will require that astronauts are well equipped with instrumentation to aid their investigations during geological field work. Two instruments are described in detail. The first is a portable X-ray Spectrometer, which can provide rapid geochemical analyses of rocks and soils, identify lunar resources and aid selection of samples for return to Earth. The second instrument is the Geological and Radiation environment package (GEORAD). This is an instrument package, mounted on a rover, to perform *in-situ* measurements on the lunar surface. It can be used for bulk geochemical measurements of rocks and soils (particularly identifying KREEP-enriched rocks), prospect for ice in shadowed areas of craters at the poles and characterise the lunar radiation environment.

## 1. Introduction

Future exploration of the Moon will involve astronauts conducting field work to address the outstanding questions in lunar geology (Crawford, 2004). Spudis and Taylor (1990) identified two types of geological investigations in a planetary context: reconnaissance and field work. Reconnaissance provides a "broad characterisation of the geological features and processes on a planetary body". This has been accomplished in the past by robotic missions, such as the unmanned Soviet Luna 21 Lunokhod rover on the floor of Le Monier crater (Florensky et al., 1978). In contrast, the objective of field work is to "understand planetary geological processes and units at all levels of detail". It is a much more open-ended activity that requires repeat visits to a site; such work demands a human presence.

The Apollo missions represent the closest approach to "true" geological field work on the Moon. The Apollo astronauts were accompanied by the Apollo Lunar Surface Experiment Package (ALSEP). This consisted of experiments that were intended operate autonomously once set up, such as a seismometer, three-axis magnetometer and thermal flux probes. In addition, the package included equipment to assist them during field work. These were predominantly sampling "tools", for example hammers, scoops, drills and core tubes (Alton, 1989). Since the time allotted for Extra Vehicular Activity (EVA) for Apollo missions was too short for a committed campaign of field work (the Apollo 17 astronauts, the last men on the Moon, spent a total of 22 h on the lunar surface), the focus was on obtaining appropriate samples for return to Earth. Even this sampling activity was restricted: during the Apollo 15 mission astronauts at Spur crater identified a geologically interesting boulder that could not be sampled since the allowed time had expired for the site (Jones, 1995).

A longer timescale for surface operations is desirable for future missions so that a detailed study of various sites surrounding the landing area can be performed (Spudis and Taylor, 1992). During field work it is preferable for rock analyses using instrumentation to be performed on the Moon as opposed to sampling followed by analysis later on Earth. Taylor and Spudis (1988) identified four reasons for this:

(i) To make *in-situ* investigations of properties of various samples likely to be disturbed by transport to a lunar habitat or Earth. For example, solar wind gases and deposits of ice in the lunar regolith are sensitive to changes in environmental parameters.
(ii) To rapidly provide data to inform further sampling and field work in the short term. Transport of rocks back to Earth for detailed analysis will be intermittent and not of the timescale feed into further operations.
(iii) To identify rocks of sufficient scientific interest to justify transport to Earth.
(iv) To support a sustained campaign of geological field work after the establishment of a permanently manned lunar base.

Instrumentation developed for robotic applications since Apollo can be adapted for use by astronauts in the field to offer a more sophisticated variety of techniques to investigate target sites. Here we describe two such instruments. The first is a portable X-ray Spectrometer to measure major and trace compositions for rocks. Secondly, we describe an instrument concept that comprises a gamma-ray spectrometer, neutron source and a particle monitor to investigate the geochemistry of rocks and soils (particularly KREEP (K – potsssium, REE – rare earth elements and P – phosphorous) enriched rocks), remotely prospect for subsurface ice deposits and monitor radiation that is harmful to humans in the lunar environment.

INSTRUMENTATION FOR GEOLOGICAL FIELD WORK ON THE MOON## 2. An X-ray Spectrometer for Lunar Field Work

2.1. Description of the X-ray spectrometer

The X-ray Spectrometer (XRS) was originally developed for the Beagle 2 mission to Mars (Sims et al., 1999). The XRS was part of the PAW package, a suite of instruments mounted on the end of the lander's robotic arm.

The scientific objectives of the XRS are to perform:

(i) A geochemical analysis of rocks and soils.
(ii) An approximate radiometric dating of rocks using the $^{40}K \to {}^{40}Ar$ technique by combining measurements of K concentration, (by the XRS) with $^{40}Ar$ concentration (by the Gas Analysis Package (Wright et al., 2003), a miniature mass spectrometer).

The Beagle 2 XRS consists of two parts: the Detector Head Assembly (DHA, shown in Figure 1), mounted on the robotic arm, and the Back End Electronics (BEE), mounted in the base of the lander. The mass of the DHA is 58 g, with a diameter of 47 mm; the BEE, consisting of a single electronics board, is 98 g and has dimensions of 120×80×15 mm. Power requirements are modest: the BEE operates from a single 6 V supply and power dissipation is no more than 4 W.

The instrument utilises excitation from radioisotope sources, identical to the XRS on the Viking landers (Clark et al., 1977), but uses the same solid state detector as used by the Alpha Proton X-ray Spectrometer on Pathfinder (Reider et al., 1997). Two $^{55}Fe$ (~60 MBq each) and two $^{109}Cd$ (~5 MBq each) provide excitation from primary X-rays of Mn K (5.90 and 6.49 keV) and Ag K (22.16 and 24.94 keV). The sources are housed in a carbon fibre reinforced plastic (CFRP) cap, the diameter of which sets the instrument's field of view at 25 mm. The fluorescent X-rays are detected by an Si-PIN diode manufactured by Amptek Inc (6 De Angelo Drive, Bedford, MA, USA). The instrument is sensitive to X-rays in the 1–27 keV range, corresponding to K-shell X-rays from elements Na to Pd.

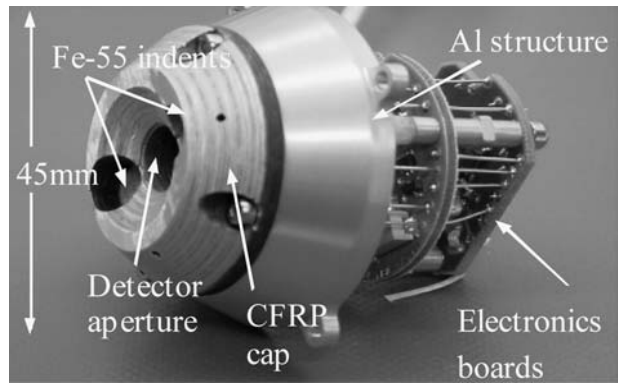

*Figure 1.* The X-ray Spectrometer Detector Head Assembly (DHA)

The Backscatter Fundamental Parameter (BFP) method (Wegrzynek et al., 2003), applicable to simultaneous $^{55}$Fe and $^{109}$Cd source excitation, has been utilised for the basis of the XRS calibration. Figure 2 is a spectrum of one of the reference materials analysed during XRS calibration (not corrected for quantum efficiency), G-2 (granite from Rhode Island).

The major elements Si, K, Ca, Ti and Fe and the trace elements Rb, Sr and Zr are clearly present in the specimen. The Rayleigh scattered Mn K$_\alpha$ and K$_\beta$ line arises from the $^{55}$Fe sources. The Ag K$_\alpha$ and K$_\beta$ Rayleigh and Compton scatter lines originate from the $^{109}$Cd sources. Other spectral artefacts include Au L$_\alpha$ lines from the wire bonds in the detector and a proportion of Ni K$_\alpha$ from the detector housing. The Fe K$_\alpha$ and Mn K$_\beta$ lines overlap, although appropriate spectral deconvolution algorithms can be used to resolve their peak areas with sufficient accuracy.

2.2. APPLICATION TO LUNAR GEOLOGICAL FIELDWORK

The low mass, power and volume of the XRS offer the possibility to repackage it for deployment not only as a terrestrial analytical instrument, but as a portable device for use by astronauts on the Moon. A portable XRS is under development for commercial purposes under the remit of a European Space Agency (ESA) Technology Transfer Program. Figure 3 shows the portable XRS during assembly.

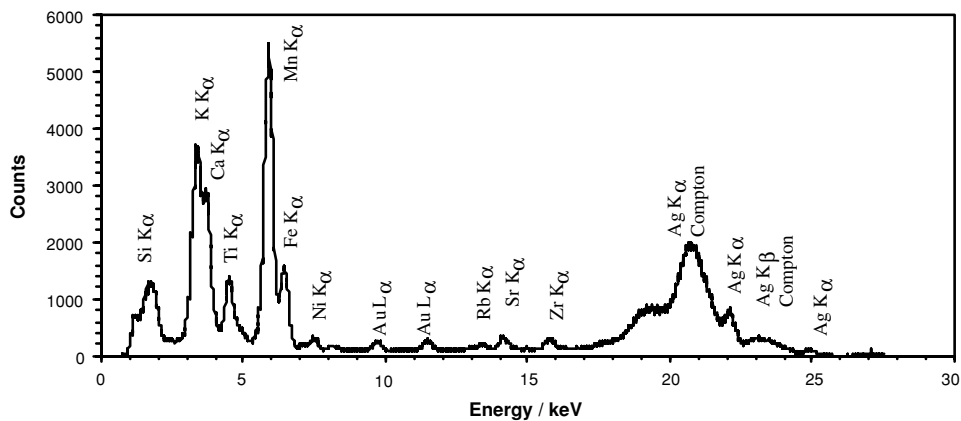

*Figure 2.* Beagle 2 XRS spectrum of granite with elemental K$_\alpha$ lines labelled. Resolution at Mn K$_\alpha$ is 331 eV.

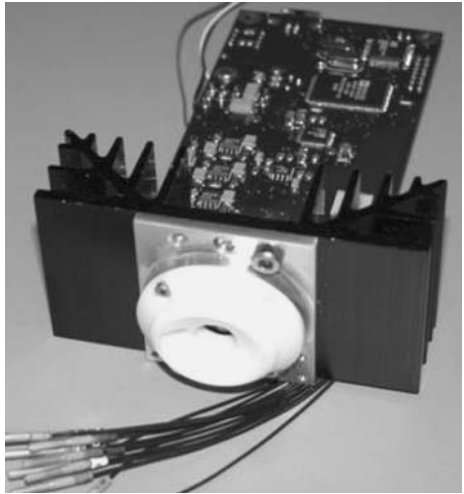

Figure 3. The portable X-ray Spectrometer during assembly.

Astronauts in the field will make excursions to different sites at which samples will be selected for return to the base for a preliminary analysis and then to Earth. Common samples must be identified and "exotic" samples distinguished for sampling since the mass allowance for return of samples to Earth is limited. The XRS can be used for rapid geochemical analyses to assist petrologic classification in the field. An astronaut in the field would place the XRS directly against, or up to approximately 5 cm away from, the target rock or soil surface and a major and trace geochemical analysis could be obtained in hundreds of seconds, depending on radioisotope source strength. The astronaut can manually perform sample preparation (eg., brushing, grinding, splitting) to investigate the geochemistry of rock interiors. Previous studies have evaluated the effects that are encountered as a result of analysing "field" geological samples using a portable X-ray spectrometer. Potts et al. (1997) developed a correction for the modification of X-ray intensities by surface roughness of rock surfaces. Okada et al. (1998) developed a correction for the effect of particle size on X-ray fluorecence intensities, although this effect can be minimised with careful design of the source-sample-detector geometry.

The XRS can also be used to identify rocks types suitable for manufacture into lunar resources. Oxygen within rocks is a valuable resource that can be utilised for rocket propellant and life support. Rodriguez (2000) evaluated feasible procedures to extract oxygen from the lunar regolith, which consequently places constraints on the rock types on the Moon that are compatible with these manufacturing processes. These include high Ti (Ilmenite bearing) basalts, pyroclastic (lunar glass) deposits and anorthosites. The Fe/Si–Al/Si and Mg/Si–Al/Si signatures of lunar rocks are a petrologic discriminator. The most useful of these to extract oxygen is Ilmenite, which can be identified geochemically from its Al/Si–Fe/Si signature. Finally, the X-ray Spectrometer can provide ground truth to measurements made by instruments that measure the geochemistry on lunar orbiters past and present such as Clementine (USA), Lunar Prospector (USA) and SMART-1 (Europe), and planned missions Chang'e I (China), Chandrayaan I (India), and SELENE (Japan).

## 3. The Geological and Radiation Environmental Monitoring Package: GEORAD

### 3.1. Description of the georad instrument suite

The GEORAD concept (Ambrosi et al., 2005) was developed in response to the Announcement of Opportunity for ESA's ExoMars rover mission (ESA, 2003) to assist in the search for extinct or extant life on Mars. It is a suite of instruments designed to be deployed on a rover. However, it can address important scientific objectives on the Moon in the context of human exploration.

The scientific objectives of the GEORAD instrument are:

(i) To determine the major (in particular K, Fe, O, Si, Al, Ca, Mg and Ti) and trace element (Th, U) composition of rocks and soils via Neutron Activation Analysis (NAA) of rocks using the gamma-ray spectrometer.
(ii) To measure the distribution of hydrogen (in the form of ice) in the subsurface using the neutron detector to detect fast (~500 keV–8 MeV) and epithermal (0–0.3 eV) neutrons and the gamma-ray spectrometer to detect the 2.2 MeV gamma-ray emission from neutron capture by hydrogen.
(iii) To characterize the radiation environment to establish the impact on the health of humans with the particle detector.

GEORAD performs geochemical investigations of rocks and soils via NAA. Interactions of ambient Galactic Cosmic Rays (GCR) with the regolith produce neutrons. The highest flux of neutrons are in the 0.5–10 MeV energy range and the peak flux is produced at approximately 0.5 m deep in the lunar regolith, although a comparable flux is produced at depths up to 2 m (Heiken et al., 1991). These neutrons are captured by atoms of elements in the regolith, which de-excite and emit gamma-rays. The energy and flux of the gamma-rays are specific to the regolith composition and are detected by the gamma-ray spectrometer. The neutron capture by hydrogen emits a 2.2 MeV gamma-ray emission which can be used to prospect for ice. In addition, a neutron source is included to enhance the background neutron signal to increase the gamma-ray flux from the regolith and therefore reduce acquisition times.

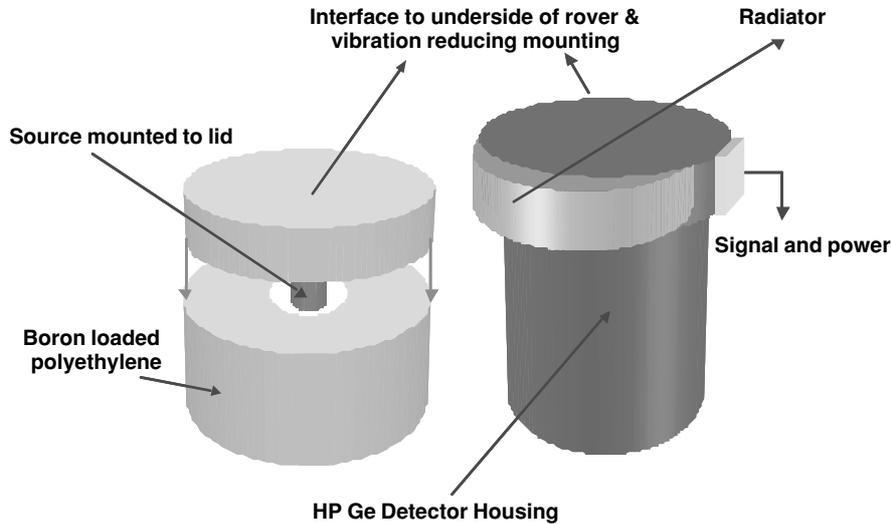

*Figure 4.* Schematic of the neutron source (left) and gamma-ray spectrometer (right).

The gamma-ray spectrometer (Figure 4, right) consists of a compound semiconductor such as cadmium telluride (CdTe), cadmium zinc telluride (CdZnTe) or a semiconductor detector such as high purity germanium (HPGe). The neutron source (Figure 4, left) consists of an alpha emitter and beryllium target such as Am–Be (also a 43 keV gamma-ray emitter) or Pu–Be provide a flux of $10^3$–$10^8$ neutrons/s into the regolith (depending upon the mass of the radioisotope source). This offers the advantage of being switchable, thereby preventing subjection of astronauts to unnecessary neutron exposure. Alternatively a pulsed deuterium–tritium (DT) tube can be used to provide a flux of $10^6$–$10^9$ neutrons/s. Meyer et al. (1995) has shown, in the case of Mars, the spectral acquisition times would be reduced by a factor of 5 by using a pulsed neutron source or radioisotopes. This time is reduced still further in the case of the Moon; the incident neutron flux to the regolith is not attenuated in the near-vacuum environment. The source emits neutrons in approximately the same energy range (the preferred sources for GEORAD are Am–Be and Pu–Be which produce neutrons in the range 4–7 MeV) as those produced from the interaction of the incident GCR flux with the regolith. Therefore, the typical penetration depths both sources of are comparable.

The solar particle and neutron backscatter detector achieves the twin objectives of prospecting for ice and monitoring the radiation environment. GEORAD can infer the presence of ice by using the neutron spectrometer to examine the backscatter of epithermal and fast neutrons in order to map the hydrogen content below the surface. To detect the charged particle flux from GCR and SPE, which is harmful to humans, it also measures the flux and energy deposited by these particles in silicon and so that an equivalent dose can be derived.

The detector consists of a stack of five fully depleted solid-state silicon detectors with the detectors mounted vertically (Figure 5). Cosmic rays penetrate these stacked detectors, the size and configuration of which means that it detects particles of energy no greater than 100 MeV. This is the desired higher energy cutoff because such particles have a low probability of interacting with human tissue. Such a detector can determine the total flux, energy distribution (with approximately a 5 MeV resolution) and directional properties of the cosmic rays. The neutrons are detected using the fibre optic scintillators coupled to the outer silicon detectors. Each scintillator is tailored to the neutron energy of interest (thermal and fast). A neutron travelling down the scintillating fibre axis will produce a light signal that is propagated down the fibres and detected by the solid-state detectors.

### 3.2. APPLICATION TO MANNED EXPLORATION OF THE MOON

GEORAD offers a sophisticated suite of instrumentation for astronauts to probe the lunar environment. The polar regions of the Moon are an attractive location for a future manned landing site or, in the longer term, a lunar base (Fristad et al., 2004). It has locations that are near permanently illuminated (Bussey et al., 2004), useful for solar power. Significantly, ice deposits potentially present could be utilised for rocket propellant and life support. A manned mission to the poles would focus on locating and exploiting these ice deposits. A GEORAD instrument could be deployed by astronauts to prospect for this ice at the poles.

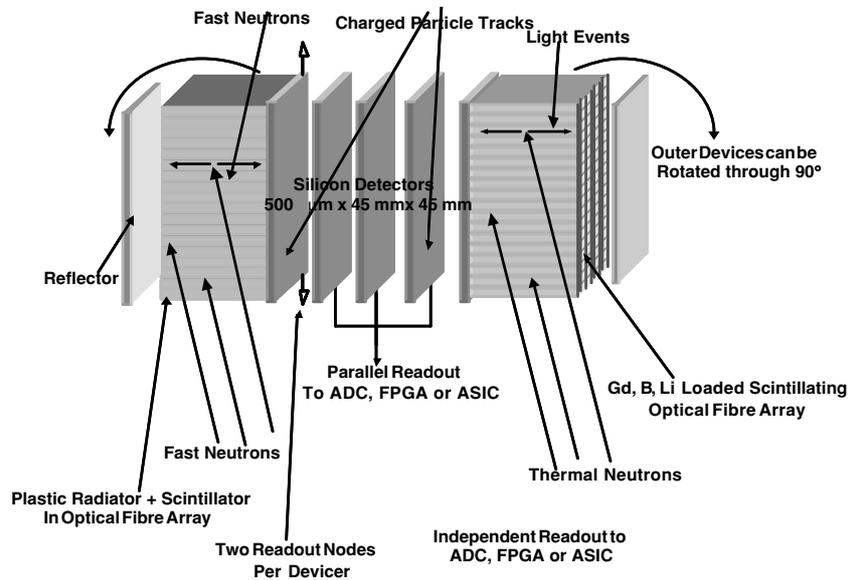

*Figure 5.* Schematic of the particle detector used to detect both charged particle flux and neutrons (thermal and fast).

Evidence for ice was found with the Lunar Prospector neutron spectrometer (Feldman et al., 1998), which detected that epithermal neutron fluxes dipped at the poles, indicating the presence of hydrogen. This data was consistent with the hydrogen in the form of water ice covered by a desiccated layer of 40 cm thick regolith within permanently shadowed craters. Suggested mechanisms for desiccation of the upper surface include sputtering by the solar wind (Lanzerotti et al., 1986), although viewed by Arnold (1979) as an insignificant effect, and incident Lyman-$\alpha$ radiation from the very local solar medium (Morgan and Shemansky, 1991). However, lunar ice delivered to the poles, from various sources suggested by Arnold (1979), can migrate to the subsurface. Cocks et al. (2002) details a mechanism whereby ice migrates to the subsurface due to temperature fluctuations and in the process is adsorbed on lunar dust. Salvail and Fanale (1994) simulated that the crater floor of Peary crater (at polar latitude of 88.6° L) has temperatures in the range 70–120 K in the shadowed region. This is a temperature compatible for ice to survive on a geological timescale.

Subsurface ice deposits can be detected by GEORAD owing to the penetration depth of neutrons and gamma-rays (order of metres) in the lunar regolith. A manned or robotic rover with a GEORAD instrument mounted on its underside could prospect for this ice. Once detected, physical sampling can be performed using a drill core. It is essential to characterise the distribution of the ice deposits *in-situ* since, once sampled, it is sensitive to changes in environmental parameters during transport.

Manned missions to the Moon are exposed to a continuous flux of Galactic Cosmic Rays (GCR) and, intermittently, intense Solar Particle Events (SPE) arising from solar flares. Had Apollo astronauts been inside the Command Module during the solar particle event that occurred in August 1972 (between the Apollo 16 and 17 missions) they would have absorbed lethal doses of radiation (4000 mSv) over a period of 10 h after the event (Hanslmeier, 2003). For comparison, the National Council on Radiation Protection and Measurement (NCRP) sets a radiation limit for astronauts in Low Earth Orbit of 250 mSv over a 30 day exposure interval (NRCP, 1989). The GEORAD particle monitor will characterise this harmful radiation on the lunar surface such that future habitats will have adequate shielding at their disposal and evaluate the risk to equipment vital for survival on the surface.

GEORAD can investigate the distribution and concentration of KREEP-enriched rocks on a local scale. KREEP elements are the concentrated residual crystallization products of the primodal lunar magma system. Thus KREEP elements act as a tracer to investigate the evolution of the crust by volcanism and major impacts. The gamma-ray spectrometer complements the X-ray spectroscopy technique since gamma-rays have a greater penetration depth in regolith (order of metres, compared to millimetres for X-ray spectroscopy). The data from the two instruments can be used to compare the uppermost layer of the regolith with the underlying volume of rock. Like the portable X-ray Spectrometer, GEORAD can provide ground truth to instruments that make similar measurements on orbiters, such as the gamma-ray and neutron spectrometer on Lunar Prospector (Feldman et al., 1999, Lawrence et al., 1998).

We have described two instruments in detail that could be utilized by astronauts during EVA to assist them during geological investigations of lunar sites. These instruments were originally developed for robotic missions, but can be adapted for use by humans on the Moon and Mars. Extended field work on the Moon by astronauts using such instrumentation could address the important science questions related to lunar origin and history. These instruments can also characterise the lunar environment to support a sustained human presence on the Moon by prospecting for lunar resources and monitoring the radiation environment.

## Acknowledgements

Talboys acknowledges support from a PPARC studentship. The development of the portable X-ray Spectrometer is funded by ESA under the remit of a Technology Transfer Programme. Finally, we would like to express thanks for the reviewer's valuable comments.

# INSTRUMENTATION FOR GEOLOGICAL FIELD WORK ON THE MOON